\documentclass[prl,twocolumn,aps,preprintnumbers,10pt]{revtex4}

\usepackage{graphicx}
\usepackage{amsmath}
\usepackage{dsfont}
\usepackage{amsfonts}
\usepackage{filecontents}
\usepackage{slashed}
\usepackage{microtype}
\usepackage[macrosonly]{chet}
\usepackage{xcolor}
\definecolor{darkblue}{rgb}{0.1,0.1,.7}
\definecolor{darkgreen}{rgb}{0,.5,0}
\usepackage[pdftex,breaklinks,colorlinks,linkcolor=darkblue,
citecolor=darkgreen]{hyperref}

\newcommand{\cO}{{\cal O}}
\newcommand{\mth}{m_{\text{th}}}
\newcommand{\fmth}{\tilde{m}_{\text{th}}}
\newcommand{\tsigma}{\tilde{\sigma}}

\DeclareMathOperator{\Disc}{Disc}
\DeclareMathOperator{\Res}{Res}
\DeclareMathOperator{\Li}{Li}
\DeclareMathOperator{\Tr}{Tr}

\begin{document}

\preprint{CERN-TH-2018-132}
\title{Dynamics of Finite-Temperature CFTs from OPE Inversion Formulas}
\author{Anastasios C. Petkou$^{1,2}$}
\author{and Andreas Stergiou$^2$}
\affiliation{$^1$Institute of Theoretical Physics, Aristotle University of Thessaloniki, Thessaloniki 54124, Greece}
\affiliation{$^2$Theoretical Physics Department, CERN, 1211 Geneva 23, Switzerland}
\date{June 2018}
\begin{abstract}
We apply the OPE inversion formula to thermal two-point functions of
bosonic and fermionic CFTs in general odd dimensions. This allows us to
analyze in detail the operator spectrum of these theories. We find that
nontrivial thermal CFTs arise when the thermal mass satisfies an algebraic
transcendental equation that ensures the absence of an infinite set of
operators from the spectrum. The solutions of these gap equations for
general odd dimensions are in general complex numbers and follow a
particular pattern. We argue that this pattern unveils the large-$N$ vacuum
structure of the corresponding theories at zero temperature.
\end{abstract}

\maketitle

\emph{Introduction.}---The description of critical systems in nontrivial
backgrounds requires data not present in the plane geometry. Perhaps the
simplest example is that of conformal field theories (CFTs) on
$S^1_{\beta}\times\mathbb{R}^{d-1}$, with $\beta$ the radius of the circle,
that describe finite-size or finite-temperature critical systems.  In such
a case, the two-point function of a scalar operator $\phi(x)$ will in
principle depend on the one-point functions of all operators that appear in
its operator product expansion (OPE) with itself, since the latter can be
nonzero. In particular, for an operator $\cO(x)$ with dimension
$\Delta_\cO$ we schematically have $\langle \cO(x)\rangle_{S^1_{\beta}
\times \mathbb{R}^{d-1}}\propto b_\cO/\beta^{\Delta_{\cO}}$, where
$b_{\cO}$ is a dimensionless parameter.

In $d=2$ the plane is conformally related to the cylinder and, although
one-point functions of conformal primaries vanish on the latter, there
exist operators such as the energy-momentum tensor which transform
anomalously under a conformal map. This fixes their one-point functions on
the cylinder and therefore the CFT data on $\mathbb{R}^2$ determine the
finite-size/finite-temperature corrections to correlation functions on
$S^1_{\beta}\times\mathbb{R}$ \cite{Cardy:1986ie, Bloete:1986qm}.

For $d>2$ there is no conformal transformation between $\mathbb{R}^d$ and
$S^1_{\beta}\times\mathbb{R}^{d-1}$, and generically one needs to find
other ways to determine the additional data $b_{\cO}$. A first step in this
direction was described in \cite{Cardy:1987dg}, where the leading
anisotropic finite-size corrections to the two-point function of scalars in
$\mathbb{R}^d$ were connected to the ratio of the thermal free-energy
density of the system and the normalization $C_T$ of the energy-momentum
tensor two-point function. An extension of these ideas to the nontrivial 3d
$O(N)$ vector model was performed in \cite{Petkou:1998fb, Petkou:1998fc},
where the relevance of the planar OPE to the description of the
finite-size/finite-temperature CFTs was demonstrated. In the latter works
the crucial point was that parameters such as $b_{\cO}$ were independently
determined by the gap equation of the vector model. In particular, once the
bosonic thermal mass was determined, all one-point functions could be
evaluated and hence the full finite-temperature two-point function could be
reconstructed.

In more recent developments, the improved understanding of CFTs on
$\mathbb{R}^d$ using numerical and analytic bootstrap methods (see
\cite{Poland:2018epd} for a recent review) calls for an extension of these
advances to finite-size/finite-temperature critical systems. In this
context an interesting work has recently appeared \cite{Iliesiu:2018fao},
whose main result is a Lorentzian inversion formula for the thermal
two-point function of a scalar $\phi(x)$ with dimension $\Delta_\phi$.
Using the OPE one can show that the Euclidean position-space \footnote{For
the corresponding momentum-space expression see \cite{Petkou:1998fb,
Petkou:1998fc}.} thermal two-point function takes the generic form
\eqn{\langle\phi(x)\phi(0)\rangle_\beta\equiv g(r,\cos\theta)=
\sum_{{\cO}_s}a_{\cO_s}\left(\frac{r}{\beta}\right)^{\Delta_{\cO_s}}
\frac{C_s^{\nu}(\cos\theta)}{r^{2\Delta_\phi}}\,,}[phiphi]
where $x^\mu=(\tau,\mathbf{x})$ are coordinates on
$S^1_{\beta}\times\mathbb{R}^{d-1}$ with period $\tau\sim\tau+\beta$,
$r=|x|$ and $\theta\in[0,\pi]$ is a polar angle when $\mathbb{R}^{d-1}$ is
written in spherical coordinates. $C_{s}^{\nu}(\cos\theta)$ are Gegenbauer
polynomials with $\nu=d/2-1$. The sum in (\ref{phiphi}) runs over all
operators ${\cO}_s$ in the OPE $\phi\times\phi$ with spin $s$ and dimension
$\Delta_{\cO_s}$. The coefficients $a_{\cO_s}$ are given by (following the
conventions of \cite{Iliesiu:2018fao})
\eqn{a_{\cO_s}=\frac{s!}{2^s(\nu)_s}
\frac{g_{\phi\phi\cO_s}b_{\cO_s}}{C_{\cO_s}}\,,}[aQs]
with $C_{\cO_s}$ and $g_{\phi\phi\cO_s}$ the corresponding two- and
three-point function coefficients, and $(a)_n$ the Pochhammer symbol. The
unit operator $\mathds{1}$ is the unique operator with dimension zero, and
here
\eqn{a_{\mathds{1}}=\frac{2^{2\Delta_\phi-d}
\Gamma(\Delta_\phi)}{\pi^{\frac{d}{2}}
\Gamma(\frac{d}{2}-\Delta_\phi)}}[aI]
so that the momentum-space two-point function is unit-normalized.

Complexifying $\Delta$ one defines the spectral function $a(\Delta,s)$ via
\eqn{g(r,\cos\theta)=\sum_s\oint_{-\epsilon-i\infty}^{-\epsilon+i\infty}\frac{d\Delta}{2\pi i}a(\Delta,s)
\frac{C_s^\nu(\cos\theta)}{r^{2\Delta_\phi-\Delta}}\,,}[aspectral]
whose poles at $\Delta=\Delta_{\cO_s}$ with residues $-a_{\cO_s}$ yield the
physical spectrum. Assuming that the physical poles lie on the right of the
imaginary axis one can close the contour clockwise for $r<1$ (we set
$\beta=1$ from now on) if $a(\Delta,s)$ does not grow exponentially at
infinity. One can then use the orthogonality of Gegenbauer polynomials (see
e.g.\ \cite[7.313]{Gradshteyn}) to project the right-hand side of
\aspectral on a spin-$s$ state and then integrate with a suitable power in
the region of convergence $r\in[0,1]$ to obtain $a(\Delta,s)$ as
\eqna{a(\Delta,s)=&\frac{1}{N_{s,\nu}}
\int_0^1\frac{dr}{r^{\Delta-2\Delta_\phi+1}}\\
&\hspace{1cm}\times
\int_{-1}^1dx\,(1-x^2)^{\nu-\frac{1}{2}}C_s^{\nu}(x)g(r,x)\,,}[Einversion]
where
\eqn{N_{s,\nu}=\frac{2^{1-2\nu}\pi\Gamma(s+2\nu)}
{(s+\nu)\Gamma(s+1)\Gamma^2(\nu)}\,.}[Nnorm]
This is termed Euclidean inversion formula in \cite{Iliesiu:2018fao}.

Writing $x=\cos\theta=(w+1/w)/2$ with $w=e^{i\theta}$ one can transform
(\ref{Einversion}) into a contour integral over the unit circle in the
complex-$w$ plane. To exploit the analytic structure of the two-point
function $g(r,\cos\theta)$ one would like to allow $w$ to explore the full
complex plane. This can be done by a suitable complexification of the
Euclidean variables $r,\theta$, defining $z=rw$ and $\bar{z}=r/w$ which are
now independent real variables.  As a function of $w$, $g(r,w)$ is assumed
to have the cuts $(-\infty,-1/r)$, $(-r,0)$, $(0,r)$ and $(1/r,\infty)$,
and to grow not faster than $w^{s_0}$ (resp.\ $1/w^{s_0}$) for large
(resp.\ small) $w$ for some constant $s_0$. Moreover, one needs to use the
analytic extension of the Gegenbauer polynomials to the whole complex plane
as
\eqn{C_{s}^{\nu}(w)=\frac{\Gamma(s+2\nu)}{\Gamma(\nu)\Gamma(s+\nu+1)}
(F_s(1/w)e^{i\nu\pi}+F_s(w)e^{-i\nu\pi})\,,}[NewGegen]
where
\eqn{F_s(w)=w^{s+2\nu}{}_2F_1(s+2\nu,\nu;s+\nu+1;w^2)\,.}[NewGegenI]
Then, the integral giving $a(\Delta,s)$ will receive contributions from the
discontinuities across the cuts of $g(r,w)$ as well as from the arcs at
infinity. The final result is
\eqn{a(\Delta,s)=a_{\text{Disc}}(\Delta,s)
+\theta(s_0-s)\,a_{\text{arcs}}(\Delta,s)\,,}[Linversion]
where
\eqna{
a_{\text{Disc}}(\Delta,s)&=K_s\int_0^1\frac{d\bar{z}}{\bar{z}}\int_1^\infty
\frac{dz}{z}\Bigl[(z\bar{z})^{\Delta_\phi-\frac{\Delta}{2}-\nu}\\
&\hspace{0.5cm}\times (z-\bar{z})^{2\nu}F_s
\left(\sqrt{\frac{\bar{z}}{z}}\right)\Disc[g(z,\bar{z})]\Bigl]\,,
}[LinversionI]
with
\eqn{
K_s=(1+(-1)^s)\frac{\Gamma(s+1)\Gamma(\nu)}{4\pi\Gamma(s+\nu)}\,.}[Knorm]
The discontinuity relevant for the evaluation of \LinversionI is the one
across the cut $(1/r,\infty)$, since all others are related to it.

\emph{Gap equations from the inversion formula.}---The OPE inversion
formulas are  powerful tools when they are applied to already known
correlation functions. In this context, one needs an ansatz for the thermal
two-point function before applying OPE inversion.  For bosons, one of the
simplest choices is to consider the momentum-space two-point function
\eqn{G^{(d)}(\omega_n,\mathbf{p})=
\frac{1}{\omega_n^2+\mathbf{p}^2+\mth^2}\,,}[prop]
where $\omega_n=2\pi n$, $n=0,\pm 1,\pm 2,\ldots,$ are the bosonic
Matsubara frequencies along the finite direction. Clearly, \prop is
motivated by known work on thermal field theory which shows that fields
develop generically a thermal mass $\mth$ at finite temperature. From our
point of view we are asking whether the simple ansatz \prop can define a
thermal CFT. We make no reference to a Lagrangian, although it is known
that \prop can be obtained, for example, in the large-$N$ limit of the
$O(N)$ model.

In arbitrary $d$ \prop can be Fourier-transformed to
\eqna{G^{(d)}(\tau,\mathbf{x})&=\frac{1}{(2\pi)^{\frac{d}{2}}}
\!\sum_{n=-\infty}^\infty\!\left(\frac{\mth}{|X_n|}\right)^{\!\frac{d}{2}-1}
\!\!\!\!K_{\frac{d}{2}-1}(\mth|X_n|)\,,\\
X_n&=(\tau-n,\mathbf{x})\,,
}[]
where $K_\alpha(x)$ is the modified Bessel function of the second kind.
Defining $z=\tau+i|\mathbf{x}|$ we have $|X_n|=\sqrt{(n-z)(n-\bar{z})}$.
From now on we focus on odd $d=2k+1$, $k=1,2,\ldots,$ and in that case we
may write~\cite[8.468]{Gradshteyn}~\footnote{In the conventions
of~\cite{Iliesiu:2018fao} the normalization $1/2^{k+1}\pi^k$ in
\eqref{propI} is rescaled away.}
\eqna{G^{(2k+1)}(\tau,\mathbf{x})&=\frac{1}{2^{k+1}\pi^k}
\sum_{n=-\infty}^\infty\frac{\mth^{k-1}}{|X_n|^k}e^{-\mth|X_n|}\\
&\hspace{2.5cm}\times \sum_{p=0}^{k-1}
\frac{L_{k,p}}{(\mth|X_n|)^p}\,,}[propI]
with
\eqn{L_{k,p}=\frac{(k-1+p)!}{2^p p!(k-1-p)!}\,.}[]
These coefficients also appear in the Bessel polynomials~\cite{Krall:1949}
\eqn{y_{n}(x)=\sum_{p=0}^{n}L_{n+1,p}\,x^p=\sqrt{\frac{2}{\pi x}}
\,e^{1/x}K_{n+\frac12}(1/x)\,.}[bessPol]

The relevant discontinuity $\Disc(G^{(d)})$ now follows simply from
understanding the discontinuity of the function
\eqn{f^{(k)}(x)=\frac{a^{k-1}}{(\sqrt{x})^k}e^{-a\sqrt{x}}
\sum_{p=0}^{k-1}\frac{L_{k,p}}{(a\sqrt{x})^p}}[]
across the cut due to the square-root branch point at $x=0$. Assuming
that the cut goes from $x=0$ to $x=\infty$ it can be verified that
\eqna{\Disc(f^{(k)}(x))&=\frac{2}{x^{k-1}}\Big(\frac{1}{\sqrt{-x}}U_k(x)\cos(a\sqrt{-x})\\
&\hspace{2cm}+V_k(x)\sin(a\sqrt{-x})\Big)\,,}[discf]
where
\eqna{U_k(x)&=\tfrac12\big(\theta_{k-1}(\sqrt{x})
  +\theta_{k-1}(-\sqrt{x})\big)\,,\\
V_k(x)&=\frac{1}{2\sqrt{x}}\big(\theta_{k-1}(\sqrt{x})
  -\theta_{k-1}(-\sqrt{x})\big)\,,
}[UVdefn]
with $\theta_{n}(x)=x^ny_n(1/x)$ the so-called reverse Bessel
polynomials~\cite{Grosswald:1978}.

Using the results \discf, \UVdefn we can now calculate \Linversion. For the
discontinuity part we find
\eqna{a_{\text{Disc}, 0}^{(k)}(\Delta,s)&=(1+(-1)^s)\frac{1}{2^{2s+k}s!}
\frac{\Gamma(k-\frac12)}{\Gamma(k+s-\frac12)}\\
&\hspace{-1.8cm}\times\!\!\!\sum_{n=0}^{k-1+s}\frac{2^{n+1}}{n!}
\frac{(2(k-1+s)-n)!}{(k-1+s-n)!}
\mth^n\Li_{2k-1+s-n}(e^{-\mth})\,,}[aNoArcs]
in the conventions of~\cite{Iliesiu:2018fao}, where
$\Li_\alpha(z)=\sum_{n=1}^\infty z^n/n^\alpha$ is the polylogarithm.  The
result \aNoArcs only pertains to the leading term in a $\bar{z}$-expansion
of the quantity under the integral in \Linversion~\footnote{This is denoted
by the subscript ``0''.}, reproducing contributions of operators with
$\Delta=d-2+s$. These are higher-spin conserved currents saturating the
unitarity bound.  Subleading terms in the $\bar{z}$-expansion can also be
considered and would lead to expressions that could be denoted by
$a_{\text{Disc},1}^{(k)}, a_{\text{Disc},2}^{(k)},\ldots,$ corresponding to
higher-twist operators.

The arc part $a_{\text{arcs}}^{(d)}(\Delta,s)$ is nonzero only for $s=0$
and in that case it needs to be taken into account carefully. We find
\eqn{a_{\text{arcs}}^{(d)}(\Delta,0)=
\frac{1}{2^{\Delta-\frac{d-5}{2}}\sqrt{\pi}}
\hspace{0.5pt}\mth^\Delta\hspace{0.5pt}\Gamma\Big(\!-\frac{\Delta}{2}\Big)
\Gamma\Big(-\frac{\Delta-d+2}{2}\Big)\,.}[aArcs]
Notice that for $\mth=0$ only the $\Delta=0$ term survives giving the
contribution of the identity operator. This, along with the corresponding
$\mth=0$ contributions from $a_{\text{Disc}}^{(k)}(\Delta,s)$, yield the
spectrum of generalized free CFTs. When $\mth\neq 0$ and for $\Delta>0$
\aArcs yields contributions of an infinite tower of scalar operators with
$\Delta=2m$, $m=1,2,\ldots,$ as well as contributions with $\Delta=d-2+2l$,
$l=0,1,2,\ldots.$ The former correspond to operators of the form
$\sigma^m$, $m=1,2,\ldots,$ where $\sigma$ is the shadow of $\phi^2$.

For the latter operators we will first focus on the $l=0$ case,
corresponding to the $\phi^2$ operator, which appears both from \aArcs and
\aNoArcs. If we demand the absence of this operator from the spectrum, as
required by the fact that it is substituted by the $\sigma$ operator, then
the residue of the $\Delta=d-2$ arc contribution should cancel the $s=0$
contribution in \aNoArcs. This turns out to give rise to a condition that
determines $\mth$, namely
\eqna{\sum_{n=0}^{k-1}\frac{2^{n+1}}{n!}\frac{(2(k-1)-n)!}{(k-1-n)!}&
\mth^n\Li_{2k-1-n}(e^{-\mth})\\
&\hspace{-0.5cm}=-\frac{1}{2\sqrt{\pi}}\hspace{0.5pt}\mth^{2k-1}
\Gamma(-k+\tfrac12)\,.}[gapEq]
This is called the gap equation and it is here presented for any $d=2k+1,
k=1,2,\ldots.$

Higher poles in \aArcs at $\Delta=d-2+2l, l=1,2,\ldots,$ correspond to
scalar operators of the form $\phi\hspace{1pt}\partial^{2l}\phi$.  Such
operators also arise from subleading terms in the $\bar{z}$ expansion of
the quantity under the integral in \Linversion, from expressions we
previously referred to as $a_{\text{Disc},1}^{(k)},
a_{\text{Disc},2}^{(k)},\ldots.$ These operators should also disappear from
the spectrum when the gap equation \gapEq is satisfied.  Although we have
verified this in a couple of cases, we do not have a general proof for it.

The arc contribution of the identity operator provides a quick consistency
check of our computations. Since the identity operator has $\Delta=0$ we
see that the pole associated with it appears due to
$\Gamma(-\frac{\Delta}{2})$ in \aArcs.  For the residue of that pole we
find
\eqn{\underset{\Delta=0}{\Res}(a_{\text{arcs}}^{(d)}(\Delta,0))
=-\frac{2^{\frac{d-3}{2}}}{\sqrt{\pi}}\Gamma(\tfrac{d}{2}-1)\,,}[]
exactly as required to reproduce the correct normalization of the identity
operator in our conventions---for this we need to take into account
$a_\mathds{1}$ from \aI and recall that we are working in conventions where
the $1/2^{k+1}\pi^k$ in \propI has been rescaled away.

It is also possible to study finite-temperature fermionic two-point
functions using the inversion formula. The simplest case  to consider is
the singlet projection of the two-point functions of Dirac fermions
$\psi_i(x)$, $\bar\psi_i(x)$ in odd dimensions,
\eqn{\langle\psi_i(x)\bar\psi_i(0)\rangle_\beta\equiv
\tilde{g}(r,\cos\theta)=\!\!\sum_{\tilde{\cO}_s\neq\mathds{1}}\!\!
\tilde{a}_{\tilde{\cO}_s}\!
\left(\frac{r}{\beta}\right)^{\Delta_{\tilde{\cO}_s}}\!
\frac{C_s^{\nu}(\cos\theta)}{r^{2\Delta_\psi}}\,,}[psibpsi]
with $\Delta_\psi=\Delta_\phi+1/2$. We denote by
$i,j=1,2,\ldots,2^{\frac{d-1}{2}}$ the spinor indices. Notice that \psibpsi
vanishes at zero temperature which means that the unit operator is absent
in the finite-temperature OPE. The corresponding unit-normalized
momentum-space two-point function is
\eqn{\tilde{G}^{(d)}(\omega_n,\mathbf{p})=
\frac{\fmth}{\omega_n^2+\mathbf{p}^2+\fmth^2}\,,}[fprop]
where the fermionic Matsubara frequencies are $\omega_n=2\pi(n+1/2)$,
$n=0,\pm1,\pm2,\ldots.$ The propagator \fprop vanishes for $\fmth=0$ so we
will only consider $\fmth\neq 0$ in the fermionic case from now on. The
calculations follow closely the bosonic case---e.g.\ it is known that
fermionic Matsubara sums reduce to a linear combination of bosonic ones. We
then notice that by virtue of the relationship $\Delta_\psi =
\Delta_\phi+1/2$, the fermionic formulas can all be obtained from the
bosonic ones by the simple shift $\Delta\to\Delta-1$. The arc contributions
in the fermionic case are thus given by
\eqna{\tilde{a}^{(d)}_{\text{arcs}}(\Delta,0)&=
-\frac{1}{2^{\Delta-\frac{d-3}{2}}
\sqrt{\pi}}\hspace{0.5pt}\fmth^{\Delta-1}\hspace{0.5pt}\Gamma
\Big(\!-\frac{\Delta-1}{2}\Big)\\
&\hspace{2.8cm}\times\Gamma\Big(\!-\frac{\Delta-d+1}{2}\Big)\,,
}[aArcsFer]
relevant for operators of dimension $\Delta=2m+1$ and $\Delta=d-1+2m$,
$m=0,1,2,\ldots.$ The former are contributions that do not arise from the
discontinuity part, having the form $\tsigma^m$ with $\tsigma$ the shadow
field of $\bar\psi\psi$. Note that, as expected, there is no contribution
from the unit operator. The latter provide contributions from operators of
the form $\bar\psi\partial^{2m}\psi$ that coincide with those coming from
the discontinuity.  The fermionic gap equation is the condition for the
cancellation of the latter operators from the spectrum and it reads
\eqna{\sum_{n=0}^{k-1}\frac{2^{n+1}}{n!}\frac{(2(k-1)-n)!}{(k-1-n)!}&
\fmth^{n+1}\Li_{2k-1-n}(-e^{-\fmth})\\
&=-\frac{1}{2\sqrt{\pi}}\hspace{0.5pt}\fmth^{2k}
\Gamma(-k+\tfrac12)\,.}[gapEqFer]

\emph{Discussion.}---One of the messages of this work is that OPE inversion
formulas can reveal the nontrivial dynamics of finite-temperature CFTs. In
the simple examples we have studied, the dynamics effect a rearrangement in
the operator spectrum which is ensured by the gap equations \gapEq and
\gapEqFer.  An analysis of the gap equations shows that their solutions
follow a pattern which, as we will argue below,  is intimately related to
the vacuum structure of scalar and fermionic theories near {\it even}
dimensions.

In the bosonic case the gap equation \gapEq in $d=3$ reads
\eqn{-\!\mth=2\log(1-e^{-\mth})\,,}[gapEqb3]
with the well-known solution
\eqn{\mth^{(d=3)}=2\log(\tfrac{1+\sqrt{5}}{2})\approx0.96242\,.}[solgapEqb3]
In $d=5$ the gap equation becomes \cite{Filothodoros:2018pdj}
\eqn{-\!\tfrac16\mth^3=\Li_3(e^{-\mth})+\mth\Li_2(e^{-\mth})\,.}[gapEqb5]
This has a complex conjugate pair of solutions given numerically by
\eqn{\mth^{(d=5)}\approx 1.17431\pm 1.19808\hspace{0.5pt}i\,.}[solmthV]
In fact, we find that for $d=3,7,11,\ldots$ the bosonic gap equation \gapEq
has a unique real solution for $\mth$ and complex solutions that come in
conjugate pairs, except in the case $d=3$ where there are no complex
solutions. To give another example, in $d=7$ we find a real and a pair of
complex conjugate solutions.  For $d=5,9,13,\ldots$ we do not find any real
solutions, and the gap equation only has pairs of complex conjugate
solutions. In $d=5$ we only find the solutions \solmthV, while in $d=9$ we
find four complex conjugate pairs of solutions. Notice also that $\mth=0$
is never a solution of the bosonic gap equations.

The fermionic gap equations in $d=3,5$ are given respectively by
\cite{Filothodoros:2018pdj}
\begin{align}-\fmth^2 &=2\fmth\log(1+e^{-\fmth})\,,\label{gapEqf3}\\
-\tfrac16\fmth^4&=\fmth\Li_3(-e^{-\fmth})+\fmth^2\Li_2(-e^{-\fmth})\label{gapEqf5}\,.
\end{align}
For $d=3$ and $\fmth\neq 0$ (\ref{gapEqf3}) has only a pair of complex
conjugate imaginary solutions $\fmth^{(d=3)}=\pm 2\pi i/3$. For $d=5$
(\ref{gapEqf5}) has a pair of opposite real solutions, as well as a pair of
complex conjugate imaginary ones which can be found numerically. This
pattern continues to higher dimensions, namely for $d=7,11,15,\ldots$ there
is no real solution to the corresponding fermionic gap equation, while for
$d=9,13,17,\ldots$ there is always a pair of opposite real solutions and an
increasing number of complex conjugate ones.

The above pattern for the solutions of bosonic and fermionic gap equations
for all \emph{odd} $d$ fits nicely with a renormalization-group
understanding of universality classes of scalars and fermions in general
dimensions.  In the bosonic case the standard lore is that the large-$N$
universality class for scalars in $d=2k+1$, $k=1,2,\ldots,$ is accessible
via the $\varepsilon$ expansion starting from $d=2k+2$. Using the
general-$d$ large-$N$ results of~\cite{Vasiliev:1981yc, Vasiliev:1981dg,
Vasiliev:1982dc} and~\cite{Lang:1990ni, Lang:1992pp, Lang:1992zw,
Petkou:1994ad, Petkou:1995vu}, this has been verified in specific cases
in~\cite{Fei:2014yja, Fei:2014xta} and~\cite{Gracey:2015xmw}. The key
ingredient in such studies is the Hubbard--Stratonovich transformation
which introduces a field $\sigma$ via the classically marginal interaction
$\sigma\phi^2$. This way $\sigma$ has dimension $\Delta_\sigma=2$ in all
$d$, and the scalars $\phi$ can be integrated out resulting in an effective
potential for $\sigma$ of the general form
\eqn{V_{\text{eff}}(\sigma) \sim \Tr_d\log
(-\partial^2+\sigma)+g_*\sigma^{\frac{d}{2}}+\cdots\,,}[Veffb]
where $g_*$ is some critical dimensionless coupling. For general $d$ the
effective potential can also receive contributions from terms involving
derivatives of $\sigma$, but the term $\sigma^{\frac{d}{2}}$ is universal.
Performing the $\Tr_d\log$ calculation in $d-\varepsilon$ one finds that
for $d=4,8,12,\ldots$ there is a resulting contribution of the form
$\sigma^{\frac{d}{2}}\log\sigma^2$, which is positive and dominates for
large $\sigma$. Thus, besides possible local minima, the effective
potential has a global minimum. On the other hand, for $d=6,10,14,\ldots$
the term $\sigma^{\frac{d}{2}}$ leads to an unbounded potential, and hence
to the absence of a global minimum, regardless of the sign of the
$\Tr_d\log$ contribution. This matches exactly the pattern we see for
$\mth$. A real $\mth$ implies a global minimum, while a complex $\mth$
signals unstable local extrema with nonzero decay width.

In the fermionic case our results are consistent with the understanding
that the corresponding large-$N$ universality classes in $d=2k+1$,
$k=1,2,\ldots$ are also accessible via the $\varepsilon$ expansion starting
from a generalization of the Gross--Neveu--Yukawa model to $d=2k+2$
\cite{ZinnJustin:1991yn}. The corresponding Hubbard--Stratonovich
transformation introduces the field $\tsigma$ via the classically marginal
interaction $\tsigma\bar{\psi}\psi$. Here $\tsigma$ has dimension
$\Delta_{\tsigma}=1$ in all $d$, and integrating out the fermions leads to
an effective potential of the form
\eqn{V_{\text{eff}}(\tsigma) \sim -\Tr_d\log (\slashed{\partial}+\tsigma)
+\tilde{g}_*\tsigma^d+\cdots\,.}[Vefff]
Notice that the $\Tr_d\log$ term enters with the opposite sign compared to
the bosonic case. In this case the universal term $\tsigma^{d}$ gives
always a bounded from below contribution (recall $d$ is even). However, the
$\Tr_d\log$ term alters the form of the effective potential in
$d-\varepsilon$. More specifically, for $d=4,8,12,\ldots$ this term gives a
negative contribution that dominates at infinity leading to an unstable
vacuum structure, while for $d=6,10,14,\ldots$ it gives a positive
contribution that guarantees the presence of a global minimum. In either
case there can be a number of unstable extrema. This matches exactly the
pattern for the $\fmth$ solutions to the fermionic gap equations.

To summarize, OPE inversion formulas applied to CFTs in nontrivial
geometries reveal crucial dynamical properties of critical systems at the
level of the operator spectrum. The consistency of the lift to the
nontrivial geometry requires that CFTs develop thermal masses that solve a
gap equation.  Remarkably, these thermal masses also encode information
about the vacuum structure of CFTs even at zero temperature.

\medskip

We would like to thank E.\ Perlmutter, J.\ Plefka, K.\ Siampos, and T.\ N.\
Tomaras for useful discussions and communications.  ACP wishes to
acknowledge the hospitality of the CERN Theory Division throughout the
completion of this work.

\bibliographystyle{apsrev4-1}
\bibliography{inversion_thermal_refs}

\end{document}